\documentclass[aps,prl,floatfix,twocolumn,tightenlines,amsmath,amssymb]{revtex4-1}

\usepackage{graphicx}
\usepackage{amssymb}
\usepackage{color}
\usepackage{psfrag}
\usepackage{ifsym}
\usepackage{epstopdf}

\newcommand{\bra}[1] {\langle #1 |}
\newcommand{\ket}[1] {| #1 \rangle}

\begin{document}
\title{Entanglement-free certification of entangling gates}

\author{M. P. Almeida$^{1,2}$, Mile Gu$^{3,4}$, Alessandro Fedrizzi$^{1,2}$, Matthew A. Broome$^{1,2}$,\\
Timothy C. Ralph$^{2}$, and Andrew G. White$^{1,2}$}
\affiliation{$^1$Centre for Engineered Quantum Systems, $^{2}$Centre for Quantum Computer and Communication Technology, School of Mathematics and Physics, University of Queensland, Brisbane,   QLD 4072, Australia\\
$^{3}$Centre for Quantum Technologies, National University of Singapore, Singapore\\
$^{4}$Center for Quantum Information, Institute for Interdisciplinary Information Sciences,
Tsinghua University, Beijing, China.}

\begin{abstract}
Not all quantum protocols require entanglement to outperform their classical alternatives. The nonclassical correlations that lead to a quantum advantage are conjectured to be captured by quantum discord. Here we demonstrate that discord has an immediate practical application: it allows a client who lacks the ability to generate entanglement or conduct quantum measurements to certify whether an untrusted party has entangling gates. We implement our protocol in the discrete-variable regime with photonic qubits, and show its success in the presence of high levels of noise and imperfect gate operations. Our technique offers a practical method to test claims of quantum processing, and for benchmarking entangling operations for physical architectures in which only highly-mixed states are available.\\
\end{abstract}

\maketitle

\bigskip
Models of intermediate quantum computing~\cite{knill1998pob,jordan2010pqc,Bremner08022011,aaronson2011ccl} offer an intriguing approach for developing quantum devices that outperform their classical counterparts. These models derive their attraction from the reduced resources compared to scalable quantum computing, and hence should be realisable sooner. One example of intermediate quantum computation is the mixed-state algorithm DQC1~\cite{knill1998pob}. Its computational advantage is often~\cite{datta2008qdp,lanyon2008eqc} associated with \emph{quantum discord}~\cite{henderson2001cqt,ollivier2001qdm}, a nonclassical correlation which is identical to entanglement for pure states, but persists for mixed states, even when the entanglement is zero.

The presence of such nonclassical correlations in virtually all mixed states prompted the question as to whether discord was ultimately a useful quantum resource~\cite{ferraro2010almost}. While it is now known that quantum circuits consisting of one- and two-qubit gates cannot provide super-polynomial computational speedups without generating discord \cite{eastin2010scc}, a formal link to computational advantage for specific protocols such as DQC1 is still missing. This has motivated extensive efforts in identifying the operational significance of discord, both in theory~\cite{oppenheim,brodutch2010quantum,piani2008no,luo2010decomposition,Boixo11,cavalcanti2011oiq,madhok2011iqd,streltsov2011lqd,piani2011anc,chuan2012rqc} and experiment~\cite{dakic2010qdr,gu2012oos}.

Here, we show that discord has an immediate practical application, the certification of entangling gates. In this scenario, Alice wishes to test whether an untrusted party, Bob, can perform entangling operations. Conventional methods requires either quantum tomography, tests of Bell inequalities, or the generation of quantum entanglement. Such actions require Alice to either conduct quantum measurements herself, possess entanglement, or put blind trust in the gate operator Bob. In many situations, this is unrealistic. Bob may represent commercial entity that markets the services of quantum processing. Alice, a potential client, would thus want to test Bob's claims with minimal technical requirements. We demonstrate this is possible when Alice can only prepare separable, but discordant, states and perform single-qubit operations. We implement our technique using a two-qubit photonic entangling gate and show that we can verify an entangling operation even in the presence of entanglement-breaking noise and imperfect gates. Note that such an asymmetry in resources is a natural assumption in adversarial quantum communication scenarios, such as \emph{blind} quantum computation~\cite{broadbent2009universal}.

We draw inspiration from the \emph{discord consumption} protocol introduced in ~\cite{gu2012oos}. In this protocol, Alice randomly encodes information in some discordant bipartite state $\rho_{AB}$, and Bob is challenged to retrieve as much of this information as possible. If Bob is limited to performing a single local measurement on each bipartition, then his performance is constrained to some incoherent limit. However, coherent bipartite interactions allow Bob to surpass this bound. The protocol suggests that discord could be used to test for Bob's capacity to coherently interact, and thus entangle the two physical systems.

Direct application of this protocol, however, leads to a loophole. The incoherent limit constrains Bob to measuring each bipartition only once. Bob can potentially cheat using multiple rounds of adaptive measurements on the two bipartitions (See Appendix A). In this letter, we close this loophole when Alice's bipartite state consists of two discordant \emph{qubits}. In this scenario, the incoherent limit strictly bounds the amount of information Bob can access with only single-qubit quantum gates. Should Bob surpass this limit, Alice can be certain that Bob has some entangling two-qubit gate.

Recall that discord quantifies the quantum component of the correlations between two physical systems \cite{henderson2001cqt,ollivier2001qdm}. The total correlations between two systems, $A$ and $B$, are quantified by the mutual information $I(A,B) {=} S(\rho_A) {+} S(\rho_B) {-} S(\rho_{AB})$, where $S(\rho)$ is the Shannon entropy of the state $\rho$. Meanwhile the classical component of these correlations, $J(A|B) {=} S(\rho_A) {-} \max_{\{\Pi_b\}\in\mathcal{M}}\sum p_b S(\rho_{A|b})$, is defined by the reduction in the entropy of $A$ after a measurement on $B$, when maximized over positive operator value measurements (POVMs) performed on $B$. (Here, $p_b$ is the probability of getting measurement outcome $b$, leaving $A$ in the conditional state $\rho_{A|b}$, $\mathcal{M}$ represents the class of all possible POVMs, and $\Pi_b$ represents a generic operator). Thus, the difference between these quantities quantifies the amount of quantum correlations between $A$ and $B$. We define this discrepancy, $\delta(A|B) {=} I(A,B) {-} J(A|B)$, as the discord. Note that discord is generally asymmetric, $\delta(A|B) \neq \delta(B|A)$.

To execute the protocol, Alice first initialises two qubits in some state $\rho_{AB}$. She then labels qubits such that $\delta(A|B) \leq \delta(B|A)$. If $\delta(A|B) \neq 0$, we say the state contains discord. Alice then generates a random variable $\mathbf{K}$ that is uniformly distributed between the four possible values $(b_1,b_2)$, where $b_1,b_2 \in \{0,1\}$ are random bits, see Fig.~1a), and encodes each possible $k =(b_1,b_2)$ on her system by application of the corresponding local unitary $U_k = \sigma_x^{b_1}\sigma_z^{b_2}$ on qubit $A$.

The qubit pair is given to Bob, who is challenged to guess $k$ by returning an estimate $k_m$ governed by a random variable $\mathbf{K}_m$. Alice quantifies Bob's performance by the amount of information $k_m$ contains about $k$, i.e., $I_{\rm{exp}} = I(\mathbf{K},\mathbf{K}_m)$, the mutual information between $\mathbf{K}$ and $\mathbf{K}_m.$

Let $I_c$ be Bob's best possible performance when he is restricted to single-qubit gates and arbitrary local measurements. Let $I_q$ be his performance when he can also implement arbitrary two-qubit gates on $A$ and $B$, or between either qubit and additional ancilla qubits: $\Delta I{=}I_q{-}I_c$ is then the `quantum advantage' of having two-qubit entangling gates. Provided $\Delta I$ is non-zero, Alice can be certain that Bob possesses some entangling two-qubit gate. In the appendix, we show that this is possible for \emph{any general} two-qubit state $\rho_{AB}$ that contains non-zero discord, i.e., $\delta(A|B) \neq 0$. Furthermore, provided $A$ and $B$ represent qubits,
\begin{equation}\label{eqn:discord_performance}
I_q - I_c = \delta(A|B).
\end{equation}
The amount of information Alice can encode within $\rho_{AB}$ that can be accessed by two-qubit operations is given exactly $\delta(A|B)$. Since virtually all mixed states contain non-zero discord, Alice has considerable freedom in her choice of $\rho_{AB}$. In practice, she will pick a state which is easy to prepare in her architecture that contains significant levels of discord.

In our proof-of-principle experiment, Alice prepares an equal mixture of the three symmetric Bell states
\begin{equation}
\rho_{AB}=\frac{1}{3}\left(\ket{\phi^{+}}\bra{\phi^{+}}+\ket{\phi^{-}}\bra{\phi^{-}}+\ket{\psi^{+}}\bra{\psi^{+}}\right),
\label{eq1}
\end{equation}
where $\ket{\phi^{\pm}}{=}(\ket{00}{\pm}\ket{11})/\sqrt{2}$ and  $\ket{\psi^{+}}{=}(\ket{01}{+}\ket{10})/\sqrt{2}$.  The state $\rho_{AB}$ can be rewritten as $\rho_{AB} {=} \sum_{i} (\ket{0_i 0_i}\bra{0_i 0_i}$ $ {+} \ket{1_i 1_i}\bra{1_i 1_i})$, where $i {=} \{x,y,z\}$, so that $\ket{0}_i$ and $\ket{1}_i$ represent the computational basis states with respect to the Pauli operators $\sigma_i$. $\rho_{AB}$ is therefore clearly separable and relatively simple to prepare: Alice can simply initialize two qubits oriented in one of the six orthogonal directions on the Bloch sphere at random. In addition, $\delta(A|B)$  has a simple form for our state of choice because $J(A|B)$ is maximized by any projective measurement. We find $\delta(A|B){=}1/3$, which ranks at the very high end of separable states \cite{qasimi2011caq}.

Bob's optimal strategy in this scenario is to conduct measurements in the Bell basis. For each $k$, the resulting state after application of $U_{k}$ will be an equal mixture of three of all four Bell states
\begin{eqnarray}
\rho_{AB} &=& \frac{1}{6} \left[\begin{array}{cccc}
2{-}b_1 & 0 & 0 & b_1 r \\
0 & 1{+}b_1 & (1{-} b_1) r & 0 \\
0 & (1{-} b_1) r & 1{+}b_1 & 0 \\
b_1 r & 0 & 0 & 2{-}b_1 \end{array}\right]
\end{eqnarray}
where $r{=}({-}1)^{b_2}$. For every instance of the protocol, Bob's Bell state measurement allows him to eliminate one of the four possible values of $k$. His probability of correctly guessing $k$ based on each outcome will be 1/3, which results in an information rate of $I_q {=} 2{-}\log_2(3) \approx 0.415$, assuming zero noise and a perfect gate operation.

%\begin{figure*}
%  \begin{center}
% \includegraphics[width=.6\textwidth]{figure_1.pdf}
%  \end{center}
%\caption{a) Quantum circuit representation of the protocol. Alice prepares $\rho_{AB}$ and encodes the classical variable $k$ by applying the unitary $U_{k}=\sigma_x^{b_1}\sigma_z^{b_2}$ to one qubit. Bob conducts a Bell state measurement to estimate Alice's encoding. b) Experimental setup. Alice's qubits are represented by the orthogonal polarisation states of two 820 nm single photons generated via type-I spontaneous parametric down-conversion in a 2~mm BBO crystal pumped by a frequency-doubled Ti:Sapphire laser (820~nm $\rightarrow$ 410~nm, 100~fs at 76~MHz). Her single-qubit rotations for preparation and encoding are implemented with quarter- (QWP) and half-wave plates (HWP). Bob's Bell state measurement consists of a non-deterministic \textsc{cz} gate, based on two-photon interference at a partially-polarising beam splitter (PPBS) with reflectivity $\eta_{V}=2/3$  ($\eta_{H}=0)$ for vertical (horizontal) polarisation, and three HWPs which enact Hadamard operations to turn the \textsc{cz} into a \textsc{cnot} gate, and to complete the Bell state measurement. Photons are detected by avalanche photodiodes (\textsc{APD}).}
%  \label{fig:figure1}
%\end{figure*}

%meaning that our protocol of testing for an entangling operation is the most general one here.
In contrast, Bob's maximal information rate \emph{without} an entangling two-qubit gate is bounded above by $I_c =I_q {-} \delta(A|B) {=} 5/3 {-} \log_2(3) {\approx} 0.082$. Upon receipt of $k_{m}$, Alice can compute Bob's achieved information rate $I_q^{\rm{exp}}$. Should this exceed $I_c$, she is sure that Bob is capable of implementing an entangling two-qubit operation.

\begin{figure}
  \begin{center}
 \includegraphics[width=\columnwidth]{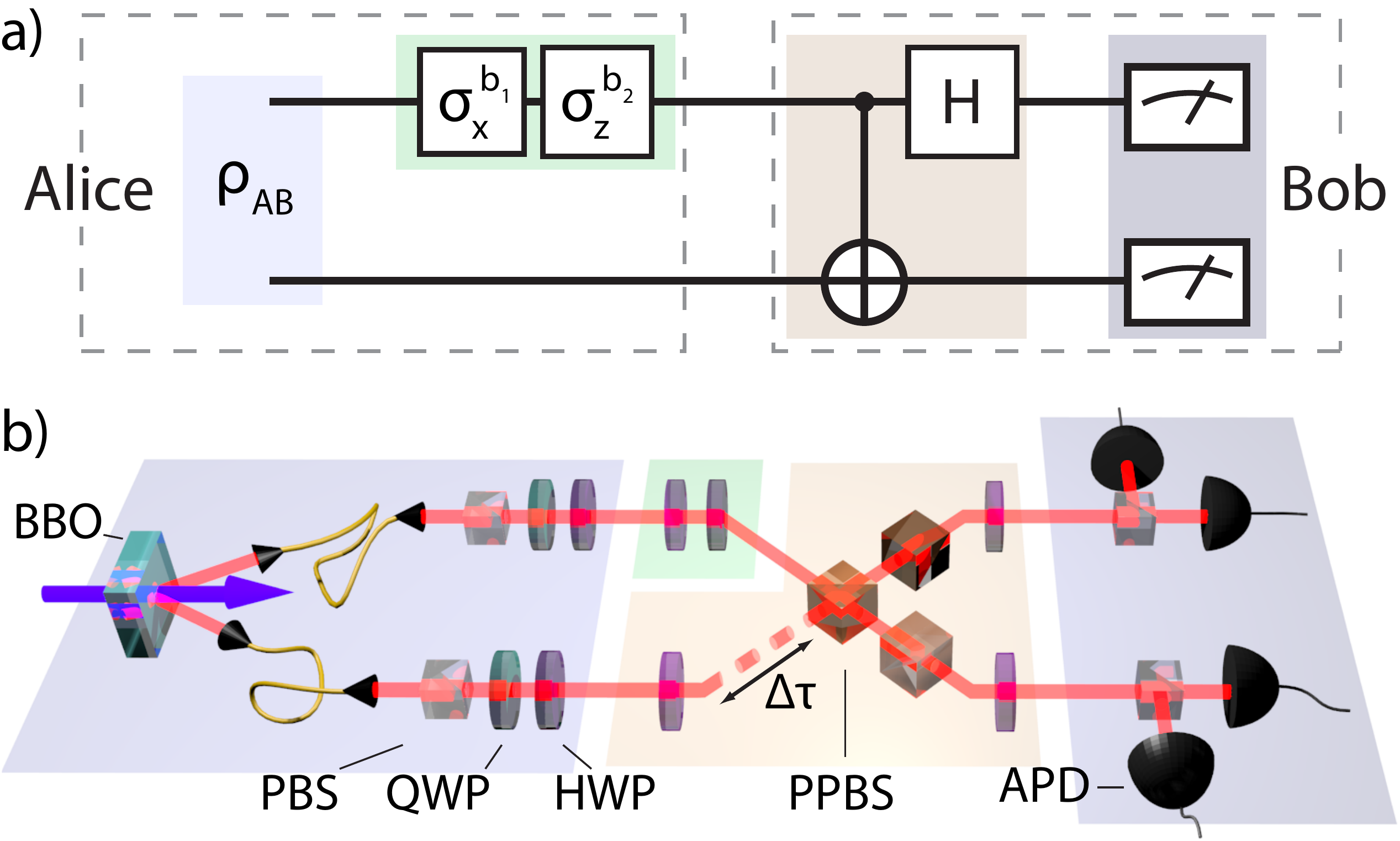}
  \end{center}
\caption{a) Quantum circuit representation of the protocol. Alice prepares discordant state $\rho_{AB}$ and encodes onto it the classical quaternary variable $k$ via the unitaries $\sigma_x^{b_1}$, $\sigma_z^{b_2}$. Bob conducts an allegedly entangling operation---optimally a Bell-state measurement---to estimate Alice's encoding. b) Experiment. Alice's qubits are realised using orthogonal polarisation states of two 820 nm single photons generated via type-I spontaneous parametric down-conversion in a 2~mm $\beta$-barium-borate (BBO) crystal pumped by a frequency-doubled (820~nm{$\rightarrow$}410~nm) Ti:Sapphire laser (100~fs pulse length, 76~MHz repetition rate). Qubits are initialised with polarising beamsplitters (PBS), and rotated (lilac area) and encoded (green area) via quarter- (QWP) and half-wave plates (HWP). Bob's entangling measurement is realised with a non-deterministic \textsc{cz} gate based on nonclassical interference of photons at a partially-polarising beam splitter (PPBS) of reflectivity $\eta_{V}=2/3$  ($\eta_{H}=0)$ for vertical (horizontal) polarisation. Photon arrival time is controlled by a relative temporal delay $\Delta \tau{=}0$, which is used to tune gate quality. The three HWPs enact Hadamard operations to turn the \textsc{cz} into a \textsc{cnot} gate, and to complete the Bell-state measurement (yellow area). Photons are analysed in the $\mathrm{Z}$-basis by PBS's, and detected by avalanche photodiodes (\textsc{APD}, grey area).}
  \label{fig:figure1}
\end{figure}

In our experiment, Alice encodes $\rho_{AB}$ in the polarisation of two single-photon qubits, where horizontal $\ket{H}$ and vertical $\ket{V}$ polarisations correspond to the logical states $\ket{0}$ and $\ket{1}$, Fig.~\ref{fig:figure1}b). Bob conducts his Bell-state measurements using a non-deterministic, controlled-phase (\textsc{cz}) gate \cite{ralph2002loc,langford2005dse} and single-qubit Hadamard gates. The \textsc{cz} gate relies on two-photon interference at a beamsplitter, imparting a $\pi$ phase shift on the input state $U_{\textsc{cz}}\ket{VV}\rightarrow -\ket{VV}$, while leaving other input combinations of basis states unchanged.

%This gate has a success probability of 1/9 which is heralded by a coincident photon detection at its two output ports.
 \begin{figure}%[hbt!]
  \begin{center}
 \includegraphics[width=.92\columnwidth]{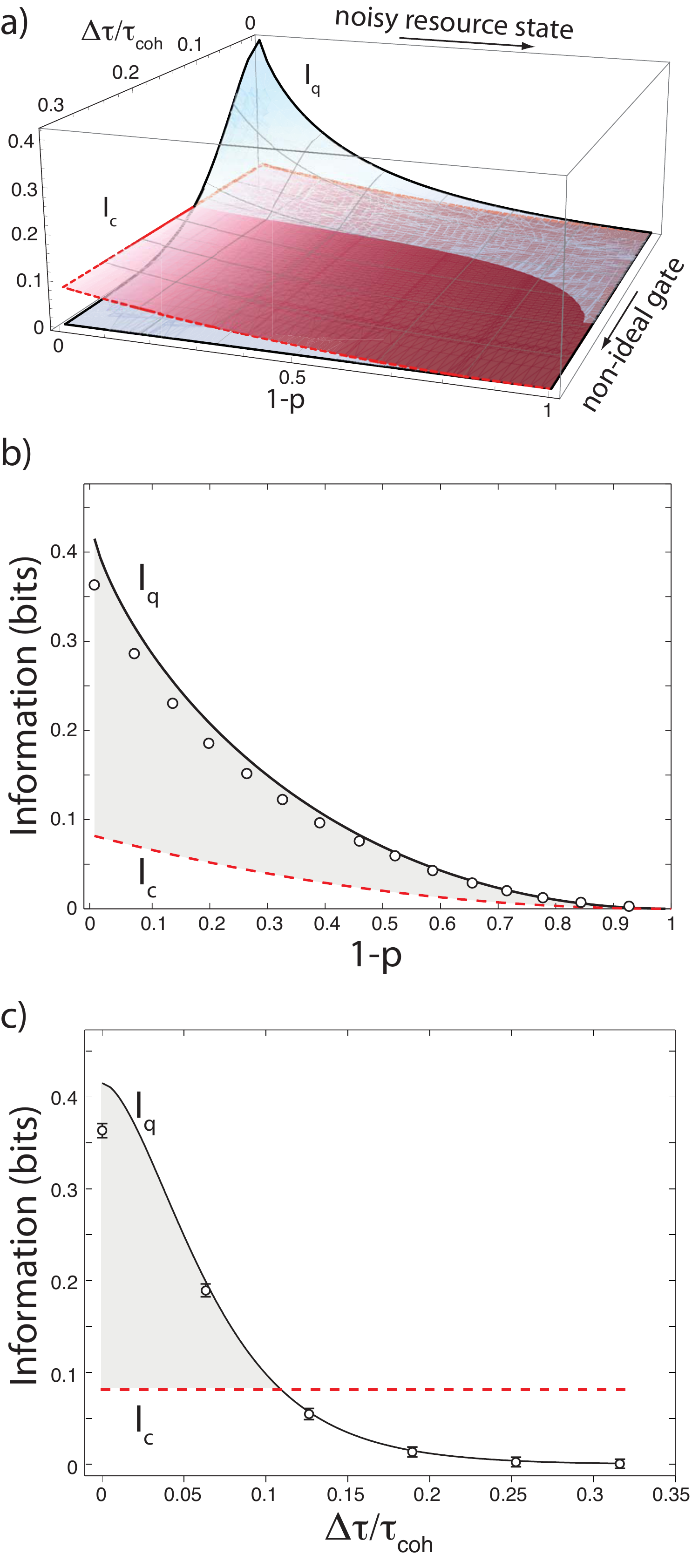}
  \end{center}
\caption{Certification of a quantum operation with discordant states. (a) Theoretical quantum performance $I_q$ achievable by Bob vs. classical limit $I_c$ as a function of white noise in Alice's resource states, $1{-}p$, and the quality of Bob's gate operation, $\Delta\tau/\tau_{coh}$. (b) Alice encodes information within noisy discordant input states $\rho(p)_{AB}=p\rho_{AB}+(1-p)\frac{\openone}{4}$. Provided Bob has access to a ideal (\textsc{cz}) gate, Bob's theoretical performance (solid black line) is guaranteed to exceed the performance limit of someone with single qubit gates (dotted red line). This quantum advantage is retain in experiment (white dots) for almost all $p$. Error bars are smaller than symbol size. (c) Indeed, even under artificial degradation of (\textsc{cz}) gate through temporal mode mismatch $\Delta\tau/\tau_{coh}$ between the interacting photonic qubits, the advantage continues to persist till $\Delta\tau/\tau_{coh}$ exceeds $0.1$. Errors are based on Poissonian counting statistics.}
  \label{fig:results}
\end{figure}

Alice constructs her discordant state $\rho_{AB}$ sequentially by preparing photons in one of the states $\{\ket{HH},\ket{VV},\ket{{D}{D}}, \ket{{A}{A}}, \ket{{R}{R}},\ket{{L}{L}}\}$, where $\ket{D},\ket{A}{=}(\ket{H}{\pm}\ket{V})/\sqrt{2}$ and $\ket{R},\ket{L}{=}(\ket{H}{\pm} i \ket{V})/\sqrt{2}$, and applies one of the four encodings $k$. Bob's Bell state measurement sums up over all components of Alice's state to extract the final measurement outcomes. The experimental information rate achieved was $I_{q}^{\textrm{exp}}{=}0.363{\pm}0.008$ which is more than 35 standard deviations above the classical limit for $I_{c}$.

We investigated the robustness of the protocol by studying two key sources of imperfection: i) the addition of white noise to the ideal state, $\rho(p)_{AB}{=}p \ \rho_{AB}{+}(1{-}p) \ \frac{\openone}{4}$; and ii) imperfect gate operation, by increasing the temporal distinguishability between the two interfering photons, $\Delta\tau$. We modeled the latter by mixing one of the \textsc{cz} gate input modes with a vacuum mode using a virtual beamsplitter with transmittivity $\xi$ \cite{ralph2002loc}: the relation of this parameter to the temporal mismatch $\Delta\tau$ is found by mapping to the well-known Gaussian two-photon interference pattern, $\xi{=}1{-}e^{{-}(\Delta\omega\Delta\tau)^2}$, where $\Delta\omega$ is the spectral bandwidth of our single photons. Starting from Bob's optimal information rate $I_{q}{\simeq}0.415$, Fig.~\ref{fig:results}(a) predicts a large operating range with quantum advantage.

We tested these predictions experimentally. In Fig.~\ref{fig:results}(b) Bob runs the entangling gate optimally, $\Delta\tau{=}0$, and Alice increases the noise on her state i.e. decreases $p$, until $\tilde{\rho}_{AB}$ is fully mixed. The ideal performance limit for Bob is in this case dictated by the Holevo limit $I_q{=}2{-}S[\rho(p)_{AB}]$. As predicted, we find that Bob \emph{always} retains a quantum advantage over the classical estimate for any given level of noise. In fact, this remains true for general noise. Any additional noise on $\rho_{AB}$ can be interpreted as initiating the protocol with some effective resource state $\rho'_{AB}$. Provided $\rho'_{AB}$ contains discord---and it generally will due to the robustness of discord to noise, Alice can use $\rho'_{AB}$ in place of $\rho_{AB}$.

%Bob's performance as a function of $p$ is given by,
%\begin{equation}
%I_q(p)=2-S(\tilde{\rho}_{AB})-\left(\frac{3+p}{4}\right)log_{2}(3).
%\label{eq4}
%\end{equation}
%Because white noise cannot increase the discord in the state Bob cannot only gain an additional information advantage \comment{Mile, is true that white noise does not increase the discord in a state?}.

In Fig.~\ref{fig:results}(c) Alice prepares the optimal state $\rho_{AB}$ and Bob decreases gate performance by temporal mode mismatch, where his optimal performance $I_q$ is now limited by the vacuum admixture, with $\xi$ taking the role of the noise parameter $p$. The amount of information $I_c$ extractable without two-qubit gates is independent of the gate operation in this scenario and therefore constant. Again, as predicted, Bob can demonstrate a quantum advantage up to ${\sim}0.1$ coherence lengths: Bob can still convince Alice he is capable of performing an entangling operation even when his gate doesn't perform very well. Conversely, if Alice knows the quality of the states she sent, she will be able to quantify the performance of Bob's entangling gate based on his guess.

%We conclude by illustrating a possible scenario where employing our technique could be particularly advantageous. For instance, there currently exist many physical architectures that promise the potential realization of a quantum computer. Some of these have achieved the capacity to perform entangling operations, but are as-of-yet unable to synthesize entanglement from such interactions due to inability of preparing quantum states of sufficient purity. While this does not necessarily preclude such processors from performing certain tasks more effectively that any classical counterpart, the lack of entanglement made it difficult to ascertain whether such architectures involve any actual quantum behavior. Our experiment demonstrates that discordant mixed-states---even quite arbitrary ones dominated by white noise---might suffice to certify that such architectures operate as advertised.
%

%The future could witness scenarios where quantum processors owned by large organizations, who sell the services of these devices to potential clients. Should a client, Alice, be approached by Bob who claims to offer such a service, she would require a method to test Bob's claims with minimal technical challenge.

Our experiment complements the recent interpretation of discord as a resource for entangling interactions. It also sheds light on the previously considered phenomenon of non locality without entanglement \cite{bennett1999qnw,pryde2005dsd}: unentangled, but discordant states can be distinguished better than zero-discord states with non-local measurements. The consequences of our protocol extend beyond the pragmatic vendor-client application we presented here. For instance, in computer science, there is significant interest in the resource asymmetry between performing a task, and verifying whether an untrusted party can perform specific computational tasks. This is reflected for example, in the study of NP problems and zero-knowledge proofs (proving to a party that you can do something without telling them how you do it). Here we show that something analogous exists for entanglement: it is possible to prove one has entangling operations without generating entanglement, provided there is some discord. It is an open question of whether this result can be generalised to n-qubit states and gates. There is no straightforward extension of our discord definition for $n$ qubits, but since 2-qubit entangling gates are universal when combined with single-qubit operations, one may bootstrap the 2-qubit certification. Meanwhile, candidate architectures for quantum computing are intrinsically entangling---such as spins in a solid interacting via J-coupling---but are often too noisy to preserve entanglement. Our technique offers an immediate method to certify whether such systems could in principle permit genuine quantum processing.

%Our experiment demonstrates that in such architectures mixed-states containing discord above the classical bound $I_c$  suffice to certify entangling operation.

%l.Some physical architectures which promise the potential realization of a quantum computer have the capacity to perform entangling operations, but are as-of-yet unable to synthesize entanglement from such interactions due to the inability to prepare sufficiently pure quantum states. While this might still allow for quantum speedups, the lack of entanglement makes it difficult to ascertain whether these architectures display actual quantum behavior. Our experiment demonstrates that mixed-states containing discord above the classical bound $I_c$  suffice to certify operation as advertised.

 \begin{acknowledgments}
We thank G. G. Gillett for help with data acquisition, K. Modi and V. Vedral for helpful discussions. This work was supported in part by the Centre for Engineered Quantum Systems (CE110001013) and the Centre for Quantum Computation and Communication Technology (CE110001027). MPA and AF acknowledge support by Australian Research Council Discovery Early Career Awards, DE120101899 and DE130100240, respectively. MG was supported by the Ministry of Education in Singapore, the National Basic Research Program of China Grant 2011CBA00300, 2011CBA00302, and the National Natural Science Foundation of China Grant 61033001, 61061130540. TCR and AGW were supported by University of Queensland Vice-Chancellor's Senior Research Fellowships.
\end{acknowledgments}

\begin{appendix}
\section{Appendix A: Proof of main result}
In this section, we prove that for an arbitrary two-qubit state $\rho_{AB}$ with discord $\delta(A|B)$, and the aforementioned encoding, Bob's advantage using entangling gates is
\begin{equation}
I_q - I_c = \delta(A|B),
\end{equation}
where $I_q$ is Bob's optimal performance with arbitrary quantum processing, and $I_c$ is the optimal performance when entangling two-qubit gates are unavailable. This is done by closing the multiple measurement loophole in~\cite{gu2012oos}.

Let $I'_c$ be Bob's optimal performance when he has no entangling gates, and furthermore, is restricted to a single measurement on each qubit. Clearly this addition restriction implies that $I'_c \leq I_c$. We will prove that additionally, $I_c \leq I'_c$, and thus $I_c = I'_c$.

This is done by contradiction. Assume that $I_c > I'_c$, i.e., Bob can exceed a performance of $I'_c$ without use of entangling gates by making multiple measurements on either qubit $A$ or qubit $B$. Let this be qubit $B$ without loss of generality.

Since $A$ resides in a 2-dimensional Hilbert space, subsequent measurements on $B$ are advantageous only if the first was weak, i.e., involving the interaction of $B$ with an ancilla $C$, followed by a measurement of $C$. This interaction, however, must have the potential to entangle $A$ and $C$ and thus constitutes an entangling gate. This contradicts out assumption that Bob did not use entangling gates. Therefore $I_c = I'_c$.

In \cite{gu2012oos}, $I'_c$ is referred to as the \emph{incoherent limit}, and it was established that
\begin{equation}\label{eqn:discordrelation}
I_q - I'_c  {=} \delta(A|B)
\end{equation}
provided Alice's choice of encoding is maximal ($\sum_k p_k U_k \rho U^k = \mathbf{I}/2$ is totally mixed for any single qubit state $\rho$). This condition is satisfied for the encoding in our protocol, thus, the relation also applies to $I_c$ and $I_q - I_c = \delta(A|B)$.

Note that if either system $A$ or $B$ is not a qubit, then the above argument does not apply, and there is potential cheating strategy for Bob. In particular, a second measurement on $B$ can still be advantageous if the first measurement on $B$ contains degeneracy. For example, if system $B$ were to consist of a composite system of two qubits, $B_1$ and $B_2$, Bob has the extra option of either measuring in the sequence, $B_1$, $A$, then $B_2$, or  $B_1$, $B_2$, then $A$, conditioned on the outcome of measuring $B_1$. Such strategies are not accounted for in the derivation of Eq. \ref{eqn:discordrelation} in \cite{gu2012oos} and in general will allow Bob to achieve a higher $I_c$.

\section{Appendix B: Explicit Evaluation of $I_c$}
Here we explicitly show that for the specific protocol where $\rho_{AB}$ is a mixture of three Bell states, Bob's  optimal performance without two-qubit gates is $I_c = \frac{5}{3} - \log_2(3)$. First, note that Bob can saturate $I_c$ by making a single $\sigma_z$ measurement on each of two qubits he receives from Alice. If the measurement results are identical, he guesses $k = (0,?)$, otherwise he guesses $k = (1,?)$, where $?$ denotes a random guess. This strategy gives no information about the second bit, but can guess the first bit correctly $2/3$ of the time. The resulting information rate is $1 - H(\frac{1}{3}) = I_c$, where $H(.)$ denotes the binary entropy.

This strategy is in fact optimal. Appendix A indicates that Bob's optimal strategy need only involve a single measurement on each qubit. Consider first a measurement on system $B$ described by operators $\{\Pi_b\}$. Since the encoding $U_k$ is localized to $A$, it commutes with the measurement operation. Therefore, if Bob were to get measurement outcome $b$, Alice would have effectively encoded onto the conditional state $\rho_{A|b}$. Bob's resulting information rate is thus constrained by the Holevo bound, $1 - \sum_b p_b S(\rho_{A|b})$, which is maximized when Bob chooses a measurement that minimizes the expected entropy of the resulting state. Due to the symmetry of $\rho_{AB}$, any projective measurement does this. Without loss of generality, measurement in the $\sigma_z$ basis gives
\begin{equation}
S(\rho_{A|b}) = \sigma_x^b\left(\frac{2}{3}\ket{0}\bra{0} + \frac{1}{3}\ket{1}\bra{1}\right) \sigma_x^b
\end{equation}
This results in a Holevo bound of $1 - H(\frac{1}{3}) = \frac{5}{3} - \log_2(3)$.

To bound the case where Bob decides to measure qubit $A$, we note that Bell states satisfy the property $(\sigma_x^{b_1}\sigma_z^{b_2} \otimes I ) \rho_{AB} (\sigma_x^{b_1}\sigma_z^{b_2} \otimes I ) =  (I \otimes \sigma_x^{b_1}\sigma_z^{b_2} ) \rho_{AB} (I \otimes \sigma_x^{b_1}\sigma_z^{b_2} )$. That is, although Alice encoded onto qubit $A$, the resulting state is functionally equivalent to encoding on qubit $B$. Thus, by inverting $A$ and $B$, the previous argument applies.

The optimal performance Bob can achieve without entangling two-qubit gates is therefore $I_c = \frac{5}{3} - \log_2(3)$. Since $\delta(A|B) = 1/3$, this agrees with our general result that $I_q - I_c  {=} \delta(A|B)$.

\end{appendix}

%
%
%\bibitem{qasimi2011caq}\bibfield  {author} {\bibinfo {author} {\bibfnamefont {A.}\ \bibnamefont
%  {Al\char21{}Qasimi}}, \bibinfo {author}, \ and {\bibfnamefont {D.~F.~V.}\ \bibnamefont {James}},\}{\bibfield  {journal} {\bibinfo  {journal} {Phys. Rev. A}\ }\textbf {\bibinfo {volume} {83}},\ \bibinfo
%  {pages} {032101} (\bibinfo {year} {2011})}


\begin{thebibliography}{26}%
\makeatletter
\providecommand \@ifxundefined [1]{%
 \@ifx{#1\undefined}
}%
\providecommand \@ifnum [1]{%
 \ifnum #1\expandafter \@firstoftwo
 \else \expandafter \@secondoftwo
 \fi
}%
\providecommand \@ifx [1]{%
 \ifx #1\expandafter \@firstoftwo
 \else \expandafter \@secondoftwo
 \fi
}%
\providecommand \natexlab [1]{#1}%
\providecommand \enquote  [1]{``#1''}%
\providecommand \bibnamefont  [1]{#1}%
\providecommand \bibfnamefont [1]{#1}%
\providecommand \citenamefont [1]{#1}%
\providecommand \href@noop [0]{\@secondoftwo}%
\providecommand \href [0]{\begingroup \@sanitize@url \@href}%
\providecommand \@href[1]{\@@startlink{#1}\@@href}%
\providecommand \@@href[1]{\endgroup#1\@@endlink}%
\providecommand \@sanitize@url [0]{\catcode `\\12\catcode `\$12\catcode
  `\&12\catcode `\#12\catcode `\^12\catcode `\_12\catcode `\%12\relax}%
\providecommand \@@startlink[1]{}%
\providecommand \@@endlink[0]{}%
\providecommand \url  [0]{\begingroup\@sanitize@url \@url }%
\providecommand \@url [1]{\endgroup\@href {#1}{\urlprefix }}%
\providecommand \urlprefix  [0]{URL }%
\providecommand \Eprint [0]{\href }%
\providecommand \doibase [0]{http://dx.doi.org/}%
\providecommand \selectlanguage [0]{\@gobble}%
\providecommand \bibinfo  [0]{\@secondoftwo}%
\providecommand \bibfield  [0]{\@secondoftwo}%
\providecommand \translation [1]{[#1]}%
\providecommand \BibitemOpen [0]{}%
\providecommand \bibitemStop [0]{}%
\providecommand \bibitemNoStop [0]{.\EOS\space}%
\providecommand \EOS [0]{\spacefactor3000\relax}%
\providecommand \BibitemShut  [1]{\csname bibitem#1\endcsname}%
\let\auto@bib@innerbib\@empty
%</preamble>
\bibitem [{\citenamefont {Knill}\ and\ \citenamefont
  {Laflamme}(1998)}]{knill1998pob}%
  \BibitemOpen
  \bibfield  {author} {\bibinfo {author} {\bibfnamefont {E.}~\bibnamefont
  {Knill}}\ and\ \bibinfo {author} {\bibfnamefont {R.}~\bibnamefont
  {Laflamme}},\ }\href@noop {} {\bibfield  {journal} {\bibinfo  {journal}
  {Phys. Rev. Lett.}\ }\textbf {\bibinfo {volume} {81}},\ \bibinfo
  {pages} {5672} (\bibinfo {year} {1998})}\BibitemShut {NoStop}%
\bibitem [{\citenamefont {Jordan}(2010)}]{jordan2010pqc}%
  \BibitemOpen
  \bibfield  {author} {\bibinfo {author} {\bibfnamefont {S.}~\bibnamefont
  {Jordan}},\ }\href@noop {} {\bibfield  {journal} {\bibinfo  {journal}
  {Quantum Information \& Computation}\ }\textbf {\bibinfo {volume} {10}},\
  \bibinfo {pages} {470} (\bibinfo {year} {2010})}\BibitemShut {NoStop}%
\bibitem [{\citenamefont {Bremner}\ \emph {et~al.}(2011)\citenamefont
  {Bremner}, \citenamefont {Jozsa},\ and\ \citenamefont
  {Shepherd}}]{Bremner08022011}%
  \BibitemOpen
  \bibfield  {author} {\bibinfo {author} {\bibfnamefont {M.~J.}\ \bibnamefont
  {Bremner}}, \bibinfo {author} {\bibfnamefont {R.}~\bibnamefont {Jozsa}}, \
  and\ \bibinfo {author} {\bibfnamefont {D.~J.}\ \bibnamefont {Shepherd}},\
  }\href@noop {} {\bibfield  {journal} {\bibinfo  {journal} {Proceedings of the
  Royal Society A: Mathematical, Physical and Engineering Science}\ }\textbf
  {\bibinfo {volume} {467}},\ \bibinfo {pages} {459} (\bibinfo {year}
  {2011})}\BibitemShut {NoStop}%
\bibitem [{\citenamefont {Aaronson}\ and\ \citenamefont
  {Arkhipov}(2011)}]{aaronson2011ccl}%
  \BibitemOpen
  \bibfield  {author} {\bibinfo {author} {\bibfnamefont {S.}~\bibnamefont
  {Aaronson}}\ and\ \bibinfo {author} {\bibfnamefont {A.}~\bibnamefont
  {Arkhipov}},\ }\href@noop {} {\bibfield  {journal} {\bibinfo  {journal}
  {Proc. ACM Symposium on Theory of Computing, San Jose, CA}\ ,\ \bibinfo
  {pages} {333}} (\bibinfo {year} {2011})}\BibitemShut {NoStop}%
\bibitem [{\citenamefont {Datta}\ \emph {et~al.}(2008)\citenamefont {Datta},
  \citenamefont {Shaji},\ and\ \citenamefont {Caves}}]{datta2008qdp}%
  \BibitemOpen
  \bibfield  {author} {\bibinfo {author} {\bibfnamefont {A.}~\bibnamefont
  {Datta}}, \bibinfo {author} {\bibfnamefont {A.}~\bibnamefont {Shaji}}, \ and\
  \bibinfo {author} {\bibfnamefont {C. M.}~\bibnamefont {Caves}},\ }\href@noop {}
  {\bibfield  {journal} {\bibinfo  {journal} {Phys. Rev. Lett.}\
  }\textbf {\bibinfo {volume} {100}},\ \bibinfo {pages} {050502} (\bibinfo
  {year} {2008})}\BibitemShut {NoStop}%
\bibitem [{\citenamefont {Lanyon}\ \emph {et~al.}(2008)\citenamefont {Lanyon},
  \citenamefont {Barbieri}, \citenamefont {Almeida},\ and\ \citenamefont
  {White}}]{lanyon2008eqc}%
  \BibitemOpen
  \bibfield  {author} {\bibinfo {author} {\bibfnamefont {B. P.}~\bibnamefont
  {Lanyon}}, \bibinfo {author} {\bibfnamefont {M.}~\bibnamefont {Barbieri}},
  \bibinfo {author} {\bibfnamefont {M. P.}~\bibnamefont {Almeida}}, \ and\
  \bibinfo {author} {\bibfnamefont {A. G.}~\bibnamefont {White}},\ }\href@noop {}
  {\bibfield  {journal} {\bibinfo  {journal} {Phys. Rev. Lett.}\
  }\textbf {\bibinfo {volume} {101}},\ \bibinfo {pages} {200501} (\bibinfo
  {year} {2008})}\BibitemShut {NoStop}%
\bibitem [{\citenamefont {Henderson}\ and\ \citenamefont
  {Vedral}(2001)}]{henderson2001cqt}%
  \BibitemOpen
  \bibfield  {author} {\bibinfo {author} {\bibfnamefont {L.}~\bibnamefont
  {Henderson}}\ and\ \bibinfo {author} {\bibfnamefont {V.}~\bibnamefont
  {Vedral}},\ }\href@noop {} {\bibfield  {journal} {\bibinfo  {journal}
  {Journal of Physics A: Mathematical and General}\ }\textbf {\bibinfo {volume}
  {34}},\ \bibinfo {pages} {6899} (\bibinfo {year} {2001})}\BibitemShut
  {NoStop}%
\bibitem [{\citenamefont {Ollivier}\ and\ \citenamefont
  {Zurek}(2001)}]{ollivier2001qdm}%
  \BibitemOpen
  \bibfield  {author} {\bibinfo {author} {\bibfnamefont {H.}~\bibnamefont
  {Ollivier}}\ and\ \bibinfo {author} {\bibfnamefont {W. H.}~\bibnamefont
  {Zurek}},\ }\href@noop {} {\bibfield  {journal} {\bibinfo  {journal}
  {Physical Review Letters}\ }\textbf {\bibinfo {volume} {88}},\ \bibinfo
  {pages} {017901} (\bibinfo {year} {2001})}\BibitemShut {NoStop}%
\bibitem [{\citenamefont {Ferraro}\ \emph {et~al.}(2010)\citenamefont
  {Ferraro}, \citenamefont {Aolita}, \citenamefont {Cavalcanti}, \citenamefont
  {Cucchietti},\ and\ \citenamefont {Acin}}]{ferraro2010almost}%
  \BibitemOpen
  \bibfield  {author} {\bibinfo {author} {\bibfnamefont {A.}~\bibnamefont
  {Ferraro}}, \bibinfo {author} {\bibfnamefont {L.}~\bibnamefont {Aolita}},
  \bibinfo {author} {\bibfnamefont {D.}~\bibnamefont {Cavalcanti}}, \bibinfo
  {author} {\bibfnamefont {F. M.}~\bibnamefont {Cucchietti}}, \ and\ \bibinfo
  {author} {\bibfnamefont {A.}~\bibnamefont {Acin}},\ }\href@noop {} {\bibfield
   {journal} {\bibinfo  {journal} {Phys. Rev. A}\ }\textbf {\bibinfo
  {volume} {81}},\ \bibinfo {pages} {052318} (\bibinfo {year}
  {2010})}\BibitemShut {NoStop}%
\bibitem [{\citenamefont {Eastin}(2010)}]{eastin2010scc}%
  \BibitemOpen
  \bibfield  {author} {\bibinfo {author} {\bibfnamefont {B.}~\bibnamefont
  {Eastin}},\ }\href@noop {} {\bibfield  {journal} {\bibinfo  {journal} {arXiv
  preprint arXiv:1006.4402}\ } (\bibinfo {year} {2010})}\BibitemShut {NoStop}%
\bibitem [{\citenamefont {Oppenheim}\ \emph {et~al.}(2002)\citenamefont
  {Oppenheim}, \citenamefont {Horodecki}, \citenamefont {Horodecki},\ and\
  \citenamefont {Horodecki}}]{oppenheim}%
  \BibitemOpen
  \bibfield  {author} {\bibinfo {author} {\bibfnamefont {J.}~\bibnamefont
  {Oppenheim}}, \bibinfo {author} {\bibfnamefont {M.}~\bibnamefont
  {Horodecki}}, \bibinfo {author} {\bibfnamefont {P.}~\bibnamefont
  {Horodecki}}, \ and\ \bibinfo {author} {\bibfnamefont {R.}~\bibnamefont
  {Horodecki}},\ }\href@noop {} {\bibfield  {journal} {\bibinfo  {journal}
  {Phys. Rev. Lett.}\ }\textbf {\bibinfo {volume} {89}},\ \bibinfo {pages}
  {180402} (\bibinfo {year} {2002})}\BibitemShut {NoStop}%
\bibitem [{\citenamefont {Brodutch}\ and\ \citenamefont
  {Terno}(2010)}]{brodutch2010quantum}%
  \BibitemOpen
  \bibfield  {author} {\bibinfo {author} {\bibfnamefont {A.}~\bibnamefont
  {Brodutch}}\ and\ \bibinfo {author} {\bibfnamefont {D. R.}~\bibnamefont
  {Terno}},\ }\href@noop {} {\bibfield  {journal} {\bibinfo  {journal}
  {Phys. Rev. A}\ }\textbf {\bibinfo {volume} {81}},\ \bibinfo {pages}
  {062103} (\bibinfo {year} {2010})}\BibitemShut {NoStop}%
\bibitem [{\citenamefont {Piani}\ \emph {et~al.}(2008)\citenamefont {Piani},
  \citenamefont {Horodecki},\ and\ \citenamefont {Horodecki}}]{piani2008no}%
  \BibitemOpen
  \bibfield  {author} {\bibinfo {author} {\bibfnamefont {M.}~\bibnamefont
  {Piani}}, \bibinfo {author} {\bibfnamefont {P.}~\bibnamefont {Horodecki}}, \
  and\ \bibinfo {author} {\bibfnamefont {R.}~\bibnamefont {Horodecki}},\
  }\href@noop {} {\bibfield  {journal} {\bibinfo  {journal} {Phys. Rev. Lett}\ }\textbf {\bibinfo {volume} {100}},\ \bibinfo {pages} {090502}
  (\bibinfo {year} {2008})}\BibitemShut {NoStop}%
\bibitem [{\citenamefont {Luo}\ and\ \citenamefont
  {Sun}(2010)}]{luo2010decomposition}%
  \BibitemOpen
  \bibfield  {author} {\bibinfo {author} {\bibfnamefont {S.}~\bibnamefont
  {Luo}}\ and\ \bibinfo {author} {\bibfnamefont {W.}~\bibnamefont {Sun}},\
  }\href@noop {} {\bibfield  {journal} {\bibinfo  {journal} {Phys. Rev.
  A}\ }\textbf {\bibinfo {volume} {82}},\ \bibinfo {pages} {012338} (\bibinfo
  {year} {2010})}\BibitemShut {NoStop}%
\bibitem [{\citenamefont {Boixo}\ \emph {et~al.}(2011)\citenamefont {Boixo},
  \citenamefont {Aolita}, \citenamefont {Cavalcanti}, \citenamefont {Modi},\
  and\ \citenamefont {Winter}}]{Boixo11}%
  \BibitemOpen
  \bibfield  {author} {\bibinfo {author} {\bibfnamefont {S.}~\bibnamefont
  {Boixo}}, \bibinfo {author} {\bibfnamefont {L.}~\bibnamefont {Aolita}},
  \bibinfo {author} {\bibfnamefont {D.}~\bibnamefont {Cavalcanti}}, \bibinfo
  {author} {\bibfnamefont {K.}~\bibnamefont {Modi}}, \ and\ \bibinfo {author}
  {\bibfnamefont {A.}~\bibnamefont {Winter}},\ }\href@noop {} {\bibfield
  {journal} {\bibinfo  {journal} {{ar{X}iv}:quant-ph/1105.2768}\ } (\bibinfo
  {year} {2011})}\BibitemShut {NoStop}%
\bibitem [{\citenamefont {Cavalcanti}\ \emph {et~al.}(2011)\citenamefont
  {Cavalcanti}, \citenamefont {Aolita}, \citenamefont {Boixo}, \citenamefont
  {Modi}, \citenamefont {Piani},\ and\ \citenamefont
  {Winter}}]{cavalcanti2011oiq}%
  \BibitemOpen
  \bibfield  {author} {\bibinfo {author} {\bibfnamefont {D.}~\bibnamefont
  {Cavalcanti}}, \bibinfo {author} {\bibfnamefont {L.}~\bibnamefont {Aolita}},
  \bibinfo {author} {\bibfnamefont {S.}~\bibnamefont {Boixo}}, \bibinfo
  {author} {\bibfnamefont {K.}~\bibnamefont {Modi}}, \bibinfo {author}
  {\bibfnamefont {M.}~\bibnamefont {Piani}}, \ and\ \bibinfo {author}
  {\bibfnamefont {A.}~\bibnamefont {Winter}},\ }\href@noop {} {\bibfield
  {journal} {\bibinfo  {journal} {Phys. Rev. A}\ }\textbf {\bibinfo
  {volume} {83}},\ \bibinfo {pages} {032324} (\bibinfo {year}
  {2011})}\BibitemShut {NoStop}%
\bibitem [{\citenamefont {Madhok}\ and\ \citenamefont
  {Datta}(2011)}]{madhok2011iqd}%
  \BibitemOpen
  \bibfield  {author} {\bibinfo {author} {\bibfnamefont {V.}~\bibnamefont
  {Madhok}}\ and\ \bibinfo {author} {\bibfnamefont {A.}~\bibnamefont {Datta}},\
  }\href@noop {} {\bibfield  {journal} {\bibinfo  {journal} {Phys. Rev.
  A}\ }\textbf {\bibinfo {volume} {83}},\ \bibinfo {pages} {032323} (\bibinfo
  {year} {2011})}\BibitemShut {NoStop}%
\bibitem [{\citenamefont {Streltsov}\ \emph {et~al.}(2011)\citenamefont
  {Streltsov}, \citenamefont {Kampermann},\ and\ \citenamefont
  {Bru{\ss}}}]{streltsov2011lqd}%
  \BibitemOpen
  \bibfield  {author} {\bibinfo {author} {\bibfnamefont {A.}~\bibnamefont
  {Streltsov}}, \bibinfo {author} {\bibfnamefont {H.}~\bibnamefont
  {Kampermann}}, \ and\ \bibinfo {author} {\bibfnamefont {D.}~\bibnamefont
  {Bru{\ss}}},\ }\href@noop {} {\bibfield  {journal} {\bibinfo  {journal}
  {Phys. Rev. Lett.}\ }\textbf {\bibinfo {volume} {106}},\ \bibinfo
  {pages} {160401} (\bibinfo {year} {2011})}\BibitemShut {NoStop}%
\bibitem [{\citenamefont {Piani}\ \emph {et~al.}(2011)\citenamefont {Piani},
  \citenamefont {Gharibian}, \citenamefont {Adesso}, \citenamefont
  {Calsamiglia}, \citenamefont {Horodecki},\ and\ \citenamefont
  {Winter}}]{piani2011anc}%
  \BibitemOpen
  \bibfield  {author} {\bibinfo {author} {\bibfnamefont {M.}~\bibnamefont
  {Piani}}, \bibinfo {author} {\bibfnamefont {S.}~\bibnamefont {Gharibian}},
  \bibinfo {author} {\bibfnamefont {G.}~\bibnamefont {Adesso}}, \bibinfo
  {author} {\bibfnamefont {J.}~\bibnamefont {Calsamiglia}}, \bibinfo {author}
  {\bibfnamefont {P.}~\bibnamefont {Horodecki}}, \ and\ \bibinfo {author}
  {\bibfnamefont {A.}~\bibnamefont {Winter}},\ }\href@noop {} {\bibfield
  {journal} {\bibinfo  {journal} {Phys. Rev. Lett.}\ }\textbf {\bibinfo
  {volume} {106}},\ \bibinfo {pages} {220403} (\bibinfo {year}
  {2011})}\BibitemShut {NoStop}%
\bibitem [{\citenamefont {Chuan}\ \emph {et~al.}(2012)\citenamefont {Chuan},
  \citenamefont {Maillard}, \citenamefont {Modi}, \citenamefont {Paterek},
  \citenamefont {Paternostro},\ and\ \citenamefont {Piani}}]{chuan2012rqc}%
  \BibitemOpen
  \bibfield  {author} {\bibinfo {author} {\bibfnamefont {T. K.}~\bibnamefont
  {Chuan}}, \bibinfo {author} {\bibfnamefont {J.}~\bibnamefont {Maillard}},
  \bibinfo {author} {\bibfnamefont {K.}~\bibnamefont {Modi}}, \bibinfo {author}
  {\bibfnamefont {T.}~\bibnamefont {Paterek}}, \bibinfo {author} {\bibfnamefont
  {M.}~\bibnamefont {Paternostro}}, \ and\ \bibinfo {author} {\bibfnamefont
  {M.}~\bibnamefont {Piani}},\ }\href@noop {} {\bibfield  {journal} {\bibinfo
  {journal} {Phys. Rev. Lett}\ }\textbf {\bibinfo {volume} {109}},\ \bibinfo
  {pages} {070501} (\bibinfo {year} {2012})}\BibitemShut {NoStop}%
\bibitem [{\citenamefont {Dakic}\ \emph {et~al.}(2012)\citenamefont {Dakic},
  \citenamefont {Lipp}, \citenamefont {Ma}, \citenamefont {Ringbauer},
  \citenamefont {Kropatschek}, \citenamefont {Barz}, \citenamefont {Paterek},
  \citenamefont {Vedral}, \citenamefont {Zeilinger}, \citenamefont {Brukner},\
  and\ \citenamefont {Walther}}]{dakic2010qdr}%
  \BibitemOpen
  \bibfield  {author} {\bibinfo {author} {\bibfnamefont {B.}~\bibnamefont
  {Dakic}}, \bibinfo {author} {\bibfnamefont {Y.~O.}\ \bibnamefont {Lipp}},
  \bibinfo {author} {\bibfnamefont {X.}~\bibnamefont {Ma}}, \bibinfo {author}
  {\bibfnamefont {M.}~\bibnamefont {Ringbauer}}, \bibinfo {author}
  {\bibfnamefont {S.}~\bibnamefont {Kropatschek}}, \bibinfo {author}
  {\bibfnamefont {S.}~\bibnamefont {Barz}}, \bibinfo {author} {\bibfnamefont
  {T.}~\bibnamefont {Paterek}}, \bibinfo {author} {\bibfnamefont
  {V.}~\bibnamefont {Vedral}}, \bibinfo {author} {\bibfnamefont
  {A.}~\bibnamefont {Zeilinger}}, \bibinfo {author} {\bibfnamefont
  {C.}~\bibnamefont {Brukner}}, \ and\ \bibinfo {author} {\bibfnamefont
  {P.}~\bibnamefont {Walther}},\ }\href@noop {} {\bibfield  {journal} {\bibinfo
   {journal} {Nat. Phys.}\ }\textbf {\bibinfo {volume} {8}},\ \bibinfo
  {pages} {666} (\bibinfo {year} {2012})}\BibitemShut {NoStop}%
\bibitem [{\citenamefont {Gu}\ \emph {et~al.}(2012)\citenamefont {Gu},
  \citenamefont {Chrzanowski}, \citenamefont {Assad}, \citenamefont {Symul},
  \citenamefont {Modi}, \citenamefont {Ralph}, \citenamefont {Vedral},\ and\
  \citenamefont {Lam}}]{gu2012oos}%
  \BibitemOpen
  \bibfield  {author} {\bibinfo {author} {\bibfnamefont {M.}~\bibnamefont
  {Gu}}, \bibinfo {author} {\bibfnamefont {H.}~\bibnamefont {Chrzanowski}},
  \bibinfo {author} {\bibfnamefont {S.}~\bibnamefont {Assad}}, \bibinfo
  {author} {\bibfnamefont {T.}~\bibnamefont {Symul}}, \bibinfo {author}
  {\bibfnamefont {K.}~\bibnamefont {Modi}}, \bibinfo {author} {\bibfnamefont
  {T.~C.}~\bibnamefont {Ralph}}, \bibinfo {author} {\bibfnamefont
  {V.}~\bibnamefont {Vedral}}, \ and\ \bibinfo {author} {\bibfnamefont
  {P.}~\bibnamefont {Lam}},\ }\href@noop {} {\bibfield  {journal} {\bibinfo
  {journal} {Nat. Phys.}\ } \textbf {\bibinfo {volume} {8}},\ \bibinfo
  {pages} {671-675} (\bibinfo {year} {2012})}\BibitemShut {NoStop}%
\bibitem [{\citenamefont {Broadbent}\ \emph {et~al.}(2009)\citenamefont {Broadbent}, \citenamefont {Fitzsimons},\ and\
  \citenamefont {Kashefi}}]{broadbent2009universal}%
    \BibitemOpen
  \bibfield  {author} {\bibinfo {author} {\bibfnamefont {A.}\ \bibnamefont
  {Broadbent}}, {\bibfnamefont {J.}\ \bibnamefont
  {Fitzsimons}} \ and\  \bibinfo {author} {\bibfnamefont {E.}\ \bibnamefont {Kashefi}},\ }
  {\bibfield  {journal} {\bibinfo
  {journal} {Foundations of Computer Science, 2009. FOCS'09. 50th Annual IEEE Symposium on}\ },\ \bibinfo
  {pages} {517--526} (\bibinfo {year} {2009})}\BibitemShut {NoStop}%
\bibitem [{\citenamefont {Al--Qasimi}\ \emph {et~al.}(2011)\citenamefont {Al--Qasimi} \ and\ \citenamefont
  {James}}]{qasimi2011caq}%
    \BibitemOpen
  \bibfield  {author} {\bibinfo {author} {\bibfnamefont {A.}\ \bibnamefont
  {Al--Qasimi}} \ and\  \bibinfo {author} {\bibfnamefont {D.~F.~V.}\ \bibnamefont {James}},\ }\href
  {\doibase 10.1103/PhysRevA.65.062324} {\bibfield  {journal} {\bibinfo
  {journal} {Phys. Rev. A}\ }\textbf {\bibinfo {volume} {83}},\ \bibinfo
  {pages} {032101} (\bibinfo {year} {2011})}\BibitemShut {NoStop}%
\bibitem [{\citenamefont {Ralph}\ \emph {et~al.}(2002)\citenamefont {Ralph},
  \citenamefont {Langford}, \citenamefont {Bell},\ and\ \citenamefont
  {White}}]{ralph2002loc}%
  \BibitemOpen
  \bibfield  {author} {\bibinfo {author} {\bibfnamefont {T.~C.}\ \bibnamefont
  {Ralph}}, \bibinfo {author} {\bibfnamefont {N.~K.}\ \bibnamefont {Langford}},
  \bibinfo {author} {\bibfnamefont {T.~B.}\ \bibnamefont {Bell}}, \ and\
  \bibinfo {author} {\bibfnamefont {A.~G.}\ \bibnamefont {White}},\ }\href
  {\doibase 10.1103/PhysRevA.65.062324} {\bibfield  {journal} {\bibinfo
  {journal} {Phys. Rev. A}\ }\textbf {\bibinfo {volume} {65}},\ \bibinfo
  {pages} {062324} (\bibinfo {year} {2002})}\BibitemShut {NoStop}%
\bibitem [{\citenamefont {Langford}\ \emph {et~al.}(2005)\citenamefont
  {Langford}, \citenamefont {Weinhold}, \citenamefont {Prevedel}, \citenamefont
  {Resch}, \citenamefont {Gilchrist}, \citenamefont {O'Brien}, \citenamefont
  {Pryde},\ and\ \citenamefont {White}}]{langford2005dse}%
  \BibitemOpen
  \bibfield  {author} {\bibinfo {author} {\bibfnamefont {N.~K.}\ \bibnamefont
  {Langford}}, \bibinfo {author} {\bibfnamefont {T.~J.}\ \bibnamefont
  {Weinhold}}, \bibinfo {author} {\bibfnamefont {R.}~\bibnamefont {Prevedel}},
  \bibinfo {author} {\bibfnamefont {K.~J.}\ \bibnamefont {Resch}}, \bibinfo
  {author} {\bibfnamefont {A.}~\bibnamefont {Gilchrist}}, \bibinfo {author}
  {\bibfnamefont {J.~L.}\ \bibnamefont {O'Brien}}, \bibinfo {author}
  {\bibfnamefont {G.~J.}\ \bibnamefont {Pryde}}, \ and\ \bibinfo {author}
  {\bibfnamefont {A.~G.}\ \bibnamefont {White}},\ }\href@noop {} {\bibfield
  {journal} {\bibinfo  {journal} {Phys. Rev. Lett.}\ }\textbf {\bibinfo
  {volume} {95}},\ \bibinfo {pages} {210504} (\bibinfo {year} {2005})}\BibitemShut
  {NoStop}%
\bibitem [{\citenamefont {Bennett}\ \emph {et~al.}(1999)\citenamefont
  {Bennett}, \citenamefont {DiVincenzo}, \citenamefont {Fuchs}, \citenamefont
  {Mor}, \citenamefont {Rains}, \citenamefont {Shor}, \citenamefont
  {Smolin},\ and\ \citenamefont {Wootters}}]{bennett1999qnw}%
  \BibitemOpen
  \bibfield  {author} {\bibinfo {author} {\bibfnamefont {C.~H.}\ \bibnamefont
  {Bennett}}, \bibinfo {author} {\bibfnamefont {D.~P.}\ \bibnamefont
  {DiVincenzo}}, \bibinfo {author} {\bibfnamefont {C.~A.}~\bibnamefont {Fuchs}},
  \bibinfo {author} {\bibfnamefont {T.}\ \bibnamefont {Mor}}, \bibinfo
  {author} {\bibfnamefont {E.}~\bibnamefont {Rains}}, \bibinfo {author}
  {\bibfnamefont {P.~W.}\ \bibnamefont {Shor}}, \bibinfo {author}
  {\bibfnamefont {J.~A.}\ \bibnamefont {Smolin}}, \ and\ \bibinfo {author}
  {\bibfnamefont {W.~K.}\ \bibnamefont {Wootters}},\ }\href@noop {} {\bibfield
  {journal} {\bibinfo  {journal} {Phys. Rev. A}\ }\textbf {\bibinfo
  {volume} {59}},\ \bibinfo {pages} {1070} (\bibinfo {year} {1999})}\BibitemShut
  {NoStop}%
\bibitem [{\citenamefont {Pryde}\ \emph {et~al.}(2005)\citenamefont
  {Pryde}, \citenamefont {O'Brien}, \citenamefont {White},\ and\ \citenamefont {Bartlett}}]{pryde2005dsd}%
  \BibitemOpen
  \bibfield  {author} {\bibinfo {author} {\bibfnamefont {G.~J.}\ \bibnamefont
  {Pryde}}, \bibinfo {author} {\bibfnamefont {J.~L.}\ \bibnamefont
  {O'Brien}}, \bibinfo {author} {\bibfnamefont {A.~G.}~\bibnamefont {White}},\ and\ \bibinfo {author}
  {\bibfnamefont {S.~D.}\ \bibnamefont {Bartlett}},\ }\href@noop {} {\bibfield
  {journal} {\bibinfo  {journal} {Phys. Rev. Lett.}\ }\textbf {\bibinfo
  {volume} {94}},\ \bibinfo {pages} {220406} (\bibinfo {year} {2005})}\BibitemShut
  {NoStop}%
   \end{thebibliography}
\end{document}